\documentclass[aps,preprint,superscriptaddress]{revtex4} 
\usepackage[T1]{fontenc}
\usepackage[dvips]{graphicx}
\usepackage[dvips]{color}
\usepackage{txfonts}
\usepackage{natbib}
\usepackage[geometry]{ifsym}

\definecolor{orange}{rgb}{1,0.6,0}
\definecolor{violet}{rgb}{0.5,0,0.5}

\begin{document}

\title{Partial decoupling between strain and polarization in mono-oriented Pb(Zr$_{0.2}$Ti$_{0.8}$)O$_3$ thin film}
\author{Pierre-Eymeric Janolin}
\affiliation{%
Laboratoire Structures, Propri\'{e}t\'{e}s et Mod\'{e}lisation des Solides, CNRS UMR 8580\\
\'{E}cole Centrale Paris, 92295 Ch\^{a}tenay-Malabry Cedex, France}
\author{Bernard Fraisse}
\affiliation{%
Laboratoire Structures, Propri\'{e}t\'{e}s et Mod\'{e}lisation des Solides, CNRS UMR 8580\\
\'{E}cole Centrale Paris, 92295 Ch\^{a}tenay-Malabry Cedex, France}
\author{Fran\c coise Le Marrec}
\affiliation{%
Laboratoire de Physique de la Mati\`{e}re Condens\'{e}e,\\
Universit\'{e} de Picardie Jules Verne, 80039 Amiens, France}
\author{Brahim Dkhil}
\affiliation{%
Laboratoire Structures, Propri\'{e}t\'{e}s et Mod\'{e}lisation des Solides, CNRS UMR 8580\\
\'{E}cole Centrale Paris, 92295 Ch\^{a}tenay-Malabry Cedex, France}
\begin{abstract}
The structural evolution of epitaxial mono-oriented (\textit{i.e.} with the \textit{c}-axis perpendicular to the interface) ferroelectric Pb(Zr$_{0.2}$,Ti$_{0.8}$)O$_3$ thin film has been investigated, using high-resolution, temperature dependent, X-ray diffraction. The full set of lattice parameters was obtained, it allowed to estimate the variation of the polarization as a function of temperature, underlying the difference between the polarization-induced tetragonality and the elastic one. The temperature evolution of the misfit strain has been calculated and found to be in good agreement with the theoretical temperature-misfit strain phase diagram. 
\end{abstract}

\maketitle
Ferroelectric materials such as Pb(Zr$_{1-x}$Ti$_x$)O$_3$ (PZT 1-x/x) solid solutions are widely investigated\cite{Frantti2002,Cox2005,Ranjan2005,Kornev2006} due to their various applications such as sensors or actuators in microelectromechanical systems (MEMS) or non volatile random access memories in storage devices. Stress is a very significant factor\cite{Sani2004,Kornev2005,Janolin2006} affecting their physical properties because of the strain-polarization coupling. Moreover, stress imposed on epitaxial PZT thin films contributes also to the drastic modification of its properties compared to its bulk form: the phase transition temperature, the phase sequence, the order of the transition may be altered. Intrinsic parameters such as the difference in lattice parameters between the bulk and the substrate, the difference in thermal properties, the spontaneous strain associated with the phase transition(s) as well as extrinsic parameters such as interface/surface and defects (oxygen vacancies and dislocations) contribute to the final strain state of the film.

However, there is only few experimental data sets showing these phase transition modifications, because of the need for high-quality mono-oriented (\textit{i.e.} with the \textit{c} axis of the film perpendicular to the interface) thin films and temperature-dependent structural characterization. Transition temperatures of ferroelectric thin films only have been compared to the one predicted by the theoretical temperature-misfit strain phase diagram\cite{Choi2004,Li2006_a, Misirlioglu2006_b,Dkhil2007,He2004,He2005_a,Haeni2004,Keane2006,Rios2006}.

In this paper, we have performed X-ray diffraction measurements to follow the temperature evolution of both in- and out-of-plane lattice parameters from room temperature up to 800K of mono-oriented epitaxial PZT 20/80 thin film. The temperature evolution of the strain is then obtained and compared to the calculated temperature-misfit strain phase diagram\cite{Pertsev2003}. In addition, the temperature evolution of the polarization has been calculated and gives a transition temperature in good agreement with the phase diagram\cite{Pertsev2003}.

100 nm thick PZT 20/80 thin film has been deposited by pulsed-laser deposition with a KrF excimer laser ($\lambda$=248 nm) on [00l]-oriented single-crystal SrTiO$_3$ substrate . The film was deposited under 0.2 mbar O$_2$ with a laser repetition rate of 2 Hz at a fixed fluence of 1.6~J/cm$^{2}$. Ceramic Pb$_{1.1}$(Zr$_{0.2}$Ti$_{0.8}$)O$_3$ target was used to deposit the 100nm film at a temperature of the substrate of 850 K. The excellent crystalline quality of the film (the rocking curve's FWHM for the (004) peak is equal to the one of the substrate) allows the determination of the thickness through finite size oscillations (not shown here).

Lattice parameters determination was carried out on a high-precision diffractometer using Cu-K$_\beta$ wavelength issued from a 18 kW rotating anode generator. The out-of-plane parameter (c$_f$) was determined from (00l)$_{l=2,3,4}$ Bragg reflections to improve accuracy and to correct any misalignment of the sample. In-plane lattice parameters (a$_f$) were determined from (204) and (024) reflections to improve accuracy assuming the in-plane lattice axis is perpendicular to the out-of-plane one. The in-plane lattice parameters have been found to be equal and perpendicular, implying a tetragonal structure from room temperature up to 800~K. The composition of the PZT film has been determined through volume calculation \cite{Frantti2002} and found to be PZT 22/78.
Fig.(\ref{fig:PZTsurSTO}) shows the temperature evolution of the lattice parameters of the PZT 20/80 film, the ones from both the target used for the deposition and the SrTiO$_3$ substrate.
	The film was found to be mono-oriented (100\% \textit{c}-domains), for every temperature monitored and no phase transition was evidenced, which suggests that the film adopts a tetragonal symmetry during deposition. This structure is in agreement with the one predicted by Alpay's domain stability map\cite{Alpay1998} (so-called "monodomain c")
The temperature-misfit strain phase diagram calculated by Pertsev \textit{et al.}\cite{Pertsev2003} is therefore appropriate to describe our film as the latter is monodomain, [00l]-oriented and deposited on a [00l]-oriented cubic substrate.

It is worth noting that the temperature dependence of the in-plane parameter of the film is parallel to the one of the substrate. Combined with a$_f$\textgreater a$_s$, it indicates that the film is strongly clamped on an \emph{effective} substrate\cite{Speck1994_a} which reflects the dislocation-modified parameter of the substrate (a$_s$*). These misfit dislocations lie at the interface between the PZT film and the substrate, releasing partially the elastic energy induced by the mismatch at the deposition temperature (T$_d$). The corresponding deposition strain $\epsilon^d$(T$_d$) describes the strain state of the film at this temperature: $\epsilon^d=\frac{a_f(T_d)-a_b^0(T_d)}{a_b^0(T_d)}$, with a$_b^0(T_d)$ the pseudocubic bulk lattice constant at T$_d$. $\epsilon^d$ is negative and the equivalent stress $\sigma^d=\frac{Y}{1-\nu}\epsilon^d$ is equal to -1.4~GPa, with Y the Young modulus and $\nu$ the Poisson ratio\footnote{We have used Y=122~GPa and $\nu$=0.36 in our calculations.}. This deposition stress is of the same order of magnitude than for PbTiO$_3$ film deposited on SrTiO$_3$\cite{Janolin2007}.
In the case of a coherent epitaxy, we would have a$_f$=a$_s$ and $\epsilon^d_{coherent}=\frac{a_s-a_b^0}{a_b^0}$, therefore the fraction of the energy released by misfit dislocations is equal to 1-$\epsilon^d/\epsilon^d_{coherent}\sim$66\%.

This description is based on the hypothesis that the film can be considered as an elastic solid which is verified as the relative volume variation is equal to the trace of the strain tensor ($\Delta V/V=tr(\epsilon)=2\epsilon_{\varparallel}+\epsilon_{\perp}=0.35\%$). Within this framework, the Poissons' coefficient may be determined from the mechanical in plane ($\epsilon_{\varparallel}$=(a$_f$-a$_b$)/a$_b$) and out-of-plane ($\epsilon_{\perp}$=(c$_f$-c$_b$)/c$_b$) strains.With a$_f$=3.966 \AA, and c$_f$=4.131 \AA, Eq.(\ref{eqnu}) gives $\nu=0.36$ for every temperature up to 800 K. 
\begin{equation}
\label{eq:epsmeca}
\epsilon_{\perp}=\frac{c_f-c_b}{c_b}=-2\frac{\nu}{1-\nu}\epsilon_{\varparallel}
\label{eqnu}
\end{equation}

Moreover, one of the hypothesis used in the calculation of the misfit-strain phase diagram is that all the dislocations are created during deposition. Our experimental data support this hypothesis. Indeed, as mentioned above, the in-plane lattice parameter of the PZT film is driven in temperature by an effective substrate. It is therefore possible to calculate the linear dislocation density, $\rho$, for dislocations lying at the interface, using Eq.(\ref{eqrho})\cite{Speck1994_b} in the compressive case, $\vert$\textbf{b}$\vert$ being the modulus of the Burger's vector :
\begin{equation}
a^*_s(T)=a_s(T)(1-\rho \vert\textbf{b}\vert)
\label{eqrho}
\end{equation}
As the temperature evolution of a$_f$(T)(=a$^*_s$(T)) and $a_s$(T) are parallel, $\rho$(T) is constant and equal to 0.4($\pm$0.1)$\cdot$10$^{6}$cm$^{-1}$. This is in good agreement with the density calculated at T$_d$ from Eq.(13) in \cite{Speck1994_b} $\rho$=0.5($\pm$0.1)$\cdot$10$^{6}$cm$^{-1}$, confirming the hypothesis that dislocations are created only during deposition.

In order to compare our experimental data with the temperature-misfit strain phase diagram, we have to calculate the misfit strain (S$^m$) at every temperature. The misfit strain is the sum of the deposition strain arising at T$_d$, which was found to be equal to -7.2$\cdot$10$^{-3}$, and of the thermal strain.
	
Between T$_d$ and room temperature (RT) a thermal strain ($\epsilon^{th}$) arises because of the difference in thermal properties between bulk PZT 20/80 and the substrate :
\begin{equation} 
\epsilon^{th}(T)=\frac{T-T_d}{a_f(RT)}\left(a_s(RT)\cdot\alpha_s-a_b^0(RT)\cdot\alpha_b\right)
\end{equation} with $\alpha_s$ and $\alpha_b$ the thermal expansion coefficient of the substrate and bulk PZT 20/80 respectively. This strain is compressive (-48 MPa) and much smaller than the deposition stress.

	We can, now, calculate the misfit strain at each temperature and plot these points on the temperature-misfit strain phase diagram proposed by \cite{Pertsev2003}. This evolution is reported on Fig.(\ref{fig:misfitPZT}). S$^m$(T)=$\epsilon^d$(T$_d$)+$\epsilon^{th}$(T) is continuous and linear, coherently with the linear evolution of a$_f$(T) and a$^0_b$(T). As $\epsilon^{th}$ and $\epsilon^d$ are negative, $\vert$S$^m\vert$ increases when temperature decreases. The predicted and observed monodomain c structures (or "\textit{c} phases") are in agreement. In addition, this evolution confirms the absence of phase transition, which implies that the film remains ferroelectric up to 800~K. Our data do not allow a direct determination of T$_C^{film}$, however a linear extrapolation on Fig.(\ref{fig:PZTsurSTO}) gives T$_C^{film}$$\sim$910$\pm$20~K. This is consistent with recent PFM measurements on a 50-nm PZT 20/80 film by Paruch and Triscone\cite{Paruch2006}.

Furthermore, it is possible to suggest a maximum temperature of the tetragonal phase (which is not necessarily ferroelectric in a thin film) from the temperature evolution of the c/a ratio (see Fig.(\ref{fig:csuraPZT})). This temperature, at which c/a=1, is around 1400~K, higher than T$_d$. The equivalent polarization can be determined from the relation c/a-1=Q.P$^2_s$ where Q is the electrostrictive coefficient. Morioka \textit{et al.} have calculated Q=0.049 m$^4$/C$^2$ for 50-nm PZT thin films (with 0.13\textless Zr/(Zr+Ti)\textless 0.65) deposited on SrTiO$_3$ with SrRuO$_3$ electrodes\cite{Morioka2004}. With this value P$_s$=90~$\mu$C/cm$^2$ in our film, which is higher than the values previously measured on tetragonal PZT films \cite{Tuttle1993,Brennecka2004}. This discrepancy comes from the fact that we have considered that the measured tetragonality was entirely due to the ferroelectricity, whereas our film is already strained at T$_d$. 
Hence, the measured c/a has to be decomposed in two parts, a purely elastic one where the strain and polarization are decoupled and a ferroelectric one where strain and polarization are coupled. Therefore c/a can be written as c/a(P)=c/a(P=0)+$\alpha$P$^2$ where c/a(P=0) and $\alpha$ corresponds, respectively, to the pure elastic strain contribution and an electrostrictive coefficient linking the strain to the polarization.

The \emph{elastic} tetragonality can be approximated considering that the tetragonality is purely elastic at T$_d$ (\textit{i.e.} we neglect the ferroelectric strain). This elastic tetragonality has to be subtracted to the measured (total) tetragonality in the calculation of the polarization (see Fig.(\ref{fig:csuraPZT})). At room temperature, the polarization determined in this way is 70~$\mu$C/cm$^2$, which is in much better agreement with previously reported values and its becomes nil at $\sim$980~K, close to the T$_C^{film}$ determined separately from the temperature-misfit strain phase diagram.

In conclusion we have deposited a mono-oriented 100-nm PZT 20/80 film on SrTiO$_3$ by PLD. The determination of the temperature evolution of the lattice parameters of the film allowed to estimate the different strain components as well as to point out that the measured tetragonality was composed of a purely elastic part and a ferroelectric one. This leads in our film to a P$_s$=70~$\mu$C/cm$^2$ at room temperature and showed the strain-polarization decoupling. Our experimental data and the theoretical misfit strain-temperature phase diagram are in very good agreement and suggest that the T$_C^{film}$ is shifted $\sim$150~K higher than in the bulk.

\newpage

	\begin{figure}[htp]
	\centering
	\caption{Lattice parameters of PZT 20/80: bulk ($\circ$) and thin film (out- \textcolor{blue}{$\blacktriangle$} and in-plane, a \textcolor{red}{$\blacktriangledown$} and b \textcolor{orange}{$\blacktriangledown$}) deposited on SrTiO$_3$ ($\vardiamondsuit$)}
	\label{fig:PZTsurSTO}
\end{figure}

	\begin{figure}[htp]
	\centering
	\caption{Misfit strain (S$^m$) as a function of temperature for our PZT 20/80 film on SrTiO$_3$ superimposed on the misfit-temperature phase diagram calculated by Pertsev \textit{et al.}\cite{Pertsev2003}. The dotted line is a guide for the eyes.}
	\label{fig:misfitPZT}
\end{figure}

\begin{figure}[htb]
	\centering
	\caption{Temperature evolutions of the tetragonality (measured $\blacktriangle$ and inferred purely ferroelectric $\blacksquare$) and polarization (calculated $\bullet$). The dotted lines represent the fits. The full line indicates the value of the purely elastic tetragonality.}
	\label{fig:csuraPZT}
\end{figure}

\newpage

\begin{figure}
	\centering
		\includegraphics{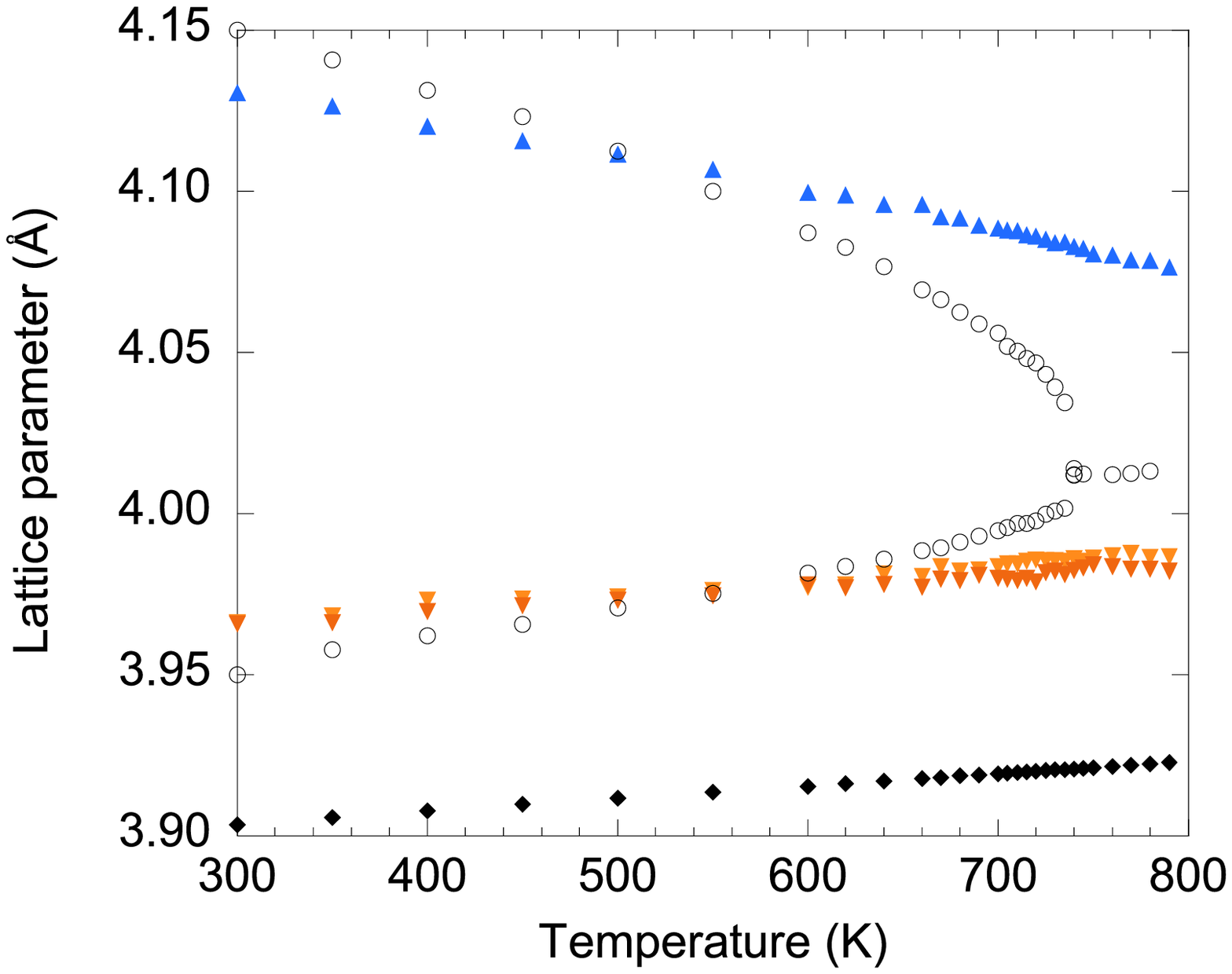}
\end{figure}

\newpage

\begin{figure}
	\centering
		\includegraphics{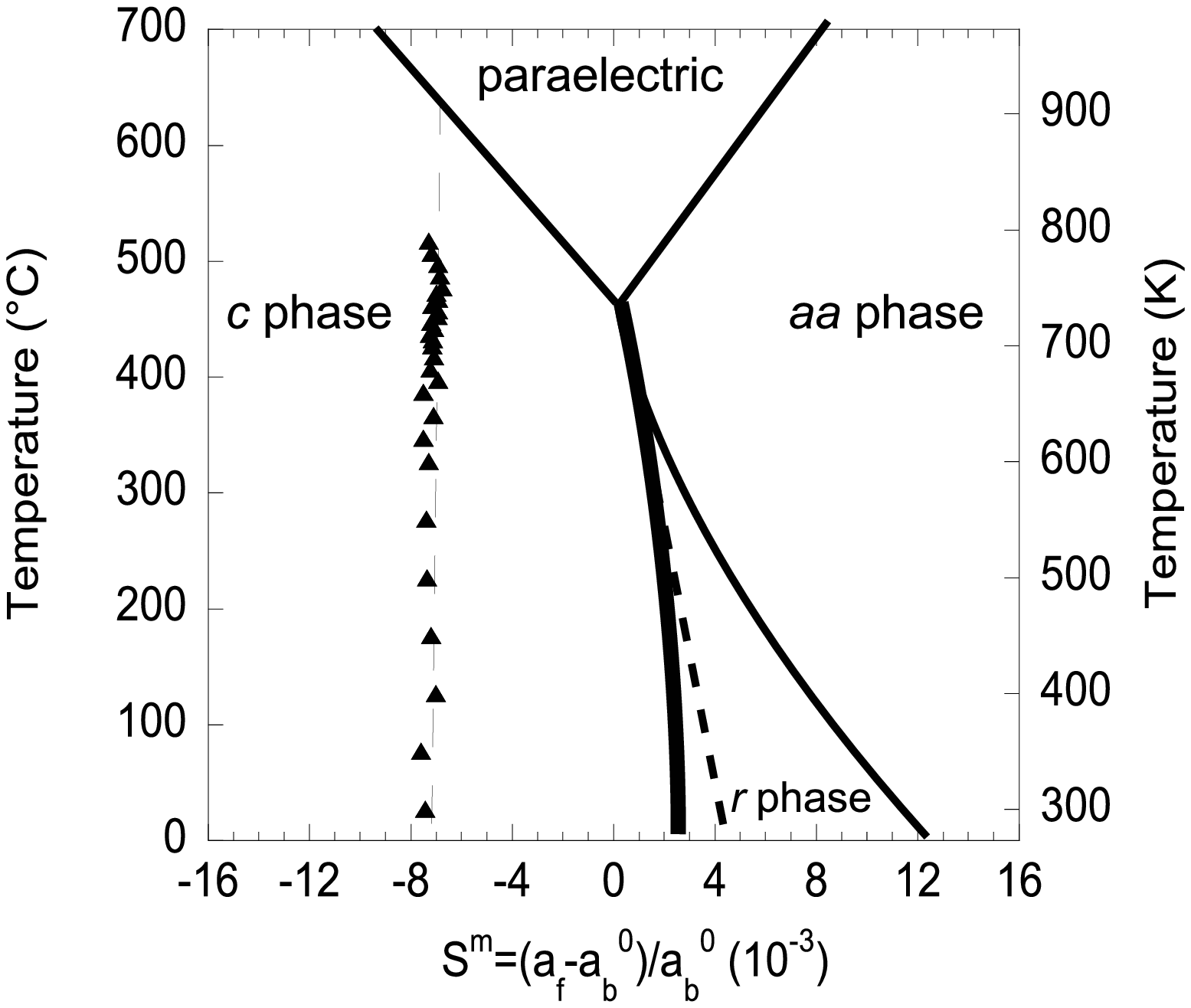}
\end{figure}

\newpage

\begin{figure}
	\centering
		\includegraphics{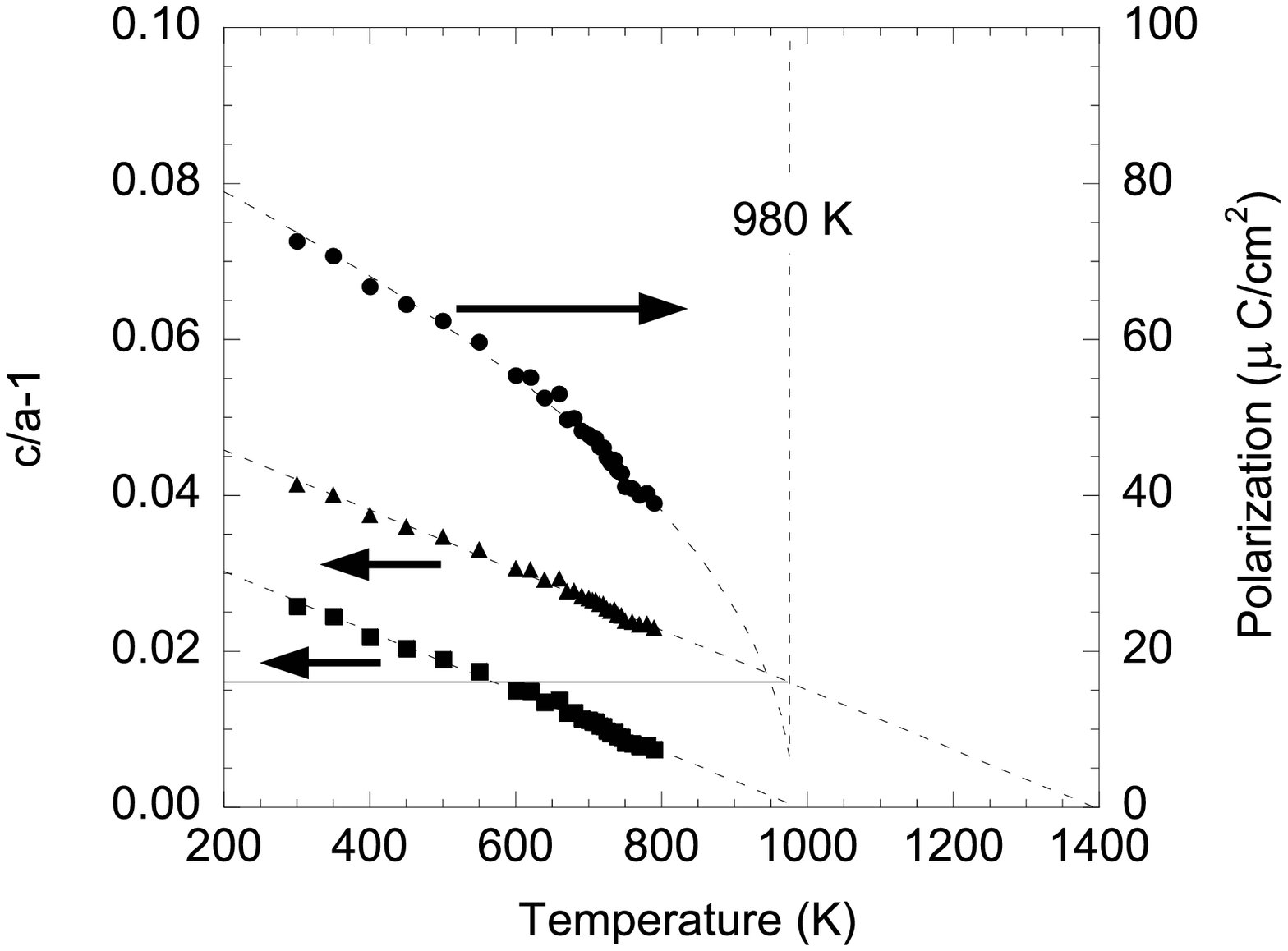}
\end{figure}

\end{document}